\pdfoutput=1
\RequirePackage{ifpdf}
\ifpdf 
\documentclass[pdftex]{sigma}
\else
\documentclass{sigma}
\fi

\numberwithin{equation}{section}

\newtheorem{Theorem}{Theorem}[section]
\newtheorem*{Theorem*}{Theorem}

\newtheorem{Conjecture}[Theorem]{Conjecture}
\newtheorem{Lemma}[Theorem]{Lemma}
\newtheorem{Proposition}[Theorem]{Proposition}

\theoremstyle{definition}
\newtheorem{Definition}[Theorem]{Definition}

\newtheorem{Remark}[Theorem]{Remark}

\newcommand\ZZ{\mathbb Z}

\renewcommand\H{\mathbb H}
\newcommand\SL{\operatorname{SL}}

\begin{document}

\allowdisplaybreaks

\newcommand{\arXivNumber}{2512.21154}

\renewcommand{\PaperNumber}{071}

\FirstPageHeading

\ShortArticleName{Limits of Equi-Affine Equi-Distant Loci of Planar Convex Domains}

\ArticleName{Limits of Equi-Affine Equi-Distant Loci of Planar\\ Convex Domains with Two Non-Parallel Asymptotes}

\Author{Nikita KALININ~$^{\rm ab}$ and Mikhail SHKOLNIKOV~$^{\rm c}$}

\AuthorNameForHeading{N.~Kalinin and M.~Shkolnikov}

\Address{$^{\rm a)}$~Guangdong Technion Israel Institute of Technology (GTIIT),\\
\hphantom{$^{\rm a)}$}~241 Daxue Road, Shantou, Guangdong Province 515603, P.R.~China}

\Address{$^{\rm b)}$~Technion-Israel Institute of Technology, Haifa, 32000, Haifa district, Israel}
\EmailD{\mail{nikaanspb@gmail.com}}

\Address{$^{\rm c)}$~Institute of Mathematics and Informatics at the Bulgarian Academy of Sciences,\\
\hphantom{$^{\rm a)}$}~Akad. G.~Bonchev St, Bl.~8, 1113 Sofia, Bulgaria}
\EmailD{\mail{m.shkolnikov@math.bas.bg}}

\ArticleDates{Received December 25, 2025, in final form July 29, 2026; Published online August 08, 2026}

\Abstract{In this note, we introduce equi-affine invariants by averaging over the space of tropical structures of fixed covolume. Applied to the tropical distance series, this construction produces a family of equi-affine invariant functions associated with convex domains which are expected to satisfy a number of remarkable properties. We conjecture a limiting description of the associated level sets in the compact case, and we prove an analogue of this statement for unbounded domains with two non-parallel asymptotes, showing the universality of the affine curvature of the resulting limiting hyperbola. In addition, we give an~explicit formula for the arithmetic mean value at the center of the unit disk.}


\Classification{53A15; 52A38; 14T90; 52A10; 11H60}

\section{Introduction}

An equi-affine plane is the affine plane $\mathbb{R}^2$ equipped with the symmetry group of area-preserving affine transformations, that is, translations together with the natural action of $\operatorname{SL}_2(\mathbb{R})$. In this geometry, ordinary Euclidean notions such as length and distance are not invariant, while area is. It is therefore natural to ask whether there exists a meaningful analogue of distance that is intrinsic to equi-affine geometry.

In this note, we introduce a family of functions that assign to each point of a convex domain a positive number and may be viewed as an equi-affine analogue of the distance to the boundary. The construction stems from a suggestion of Conan Leung who proposed to average the tropical distance function over the space of all tropical structures of fixed co-area of the ambient plane. A~guiding expectation is that this averaging procedure is related to classical affine-geometric objects such as the Monge--Amp\`ere equation and affine normal flow. See a popular exposition on how the optimal transport problem leads to the Monge--Amp\'ere equation \cite{caffarelli2004monge} and a polyhedral approximation for its solution by Pogorelov \cite{gutierrez2016monge, pogorelov1964monge,pogorelov1971regularity, pogorelov1975minkowski,1988}.

\begin{figure}[!ht] \centering

\includegraphics[width=0.5\linewidth]{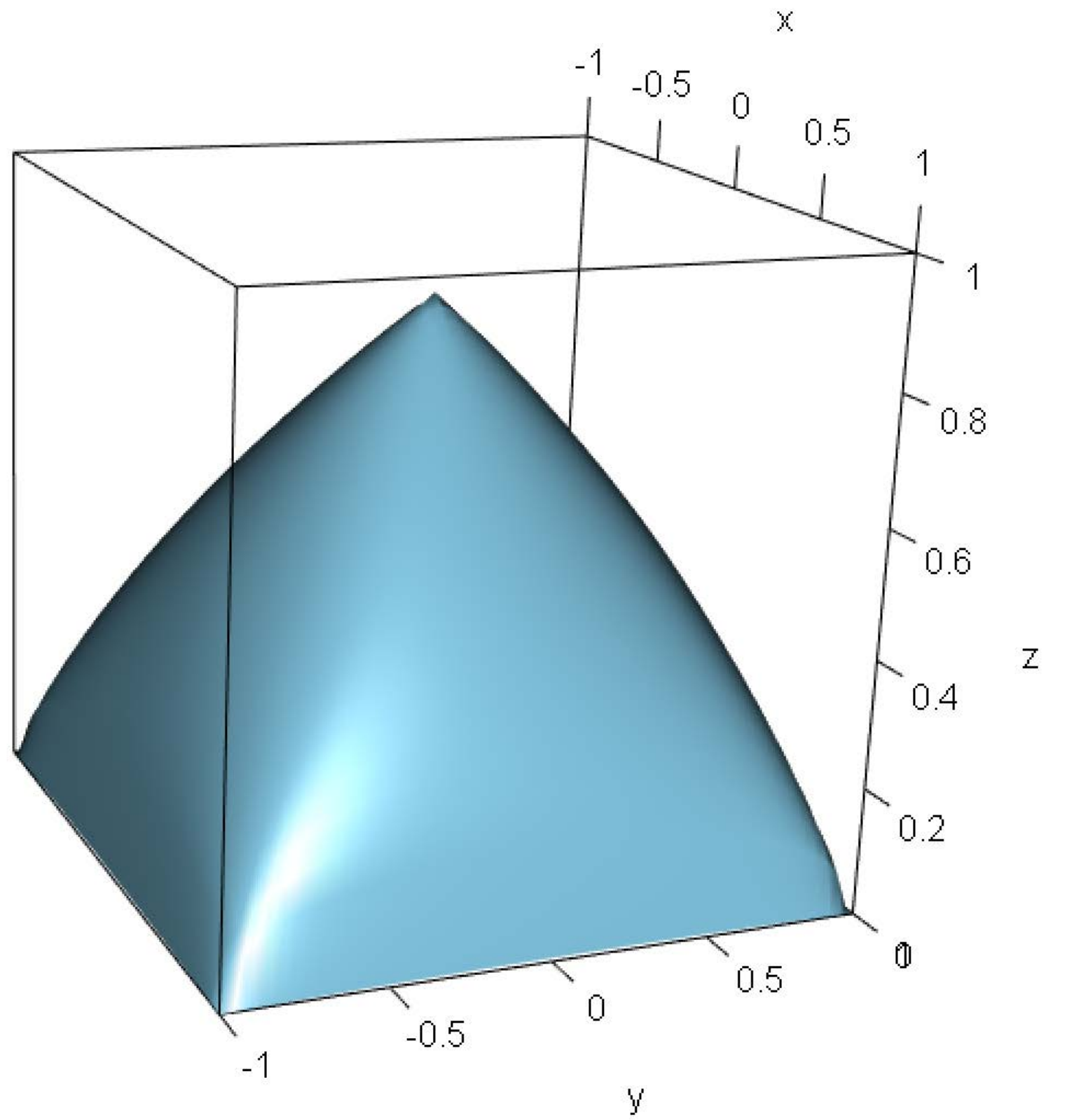}

\caption{A plot of the arithmetic average $\mathcal{A}^{\square}_1$ of tropical distance over coarea one tropical structures for the square $\square=[-1,1]^2$. Observe the conical singularity in the middle, as well as the apparent smoothness in all other points. The level sets are shown on Figure~\ref{fig:levels}.}\label{fig_square}
 \end{figure}

Remarkably, in contrast with the tropical distance function, whose non-zero levels are polygonal, equi-affine distance function seems to have smooth level sets, which we may call ``equi-distant loci'' of a domain, even if we start with a polygon (Figure \ref{fig_square} shows a plot on a square). Our purpose is to initiate the study of these level sets. A particular conjecture that we put forward is that for a fixed compact convex domain, these loci, after an appropriate rescaling, converge to an ellipse. In fact, as a few numerical simulations performed by Ernesto Lupercio suggest, these level sets have monotonically increasing Mahler area (which is a scale free affine invariant roughly measuring the ``roundness'' of a domain) approaching that of the global maximum in the space of convex domains, i.e., that of ellipses. Thus, one can expect that developing the theory in higher dimensions may give a tool to tackle Mahler conjecture~\cite{mahler}, which is a fundamental unsolved problem in affine and convex geometries.

 \begin{figure}[!ht]
 \centering
 \includegraphics[width=0.5\linewidth]{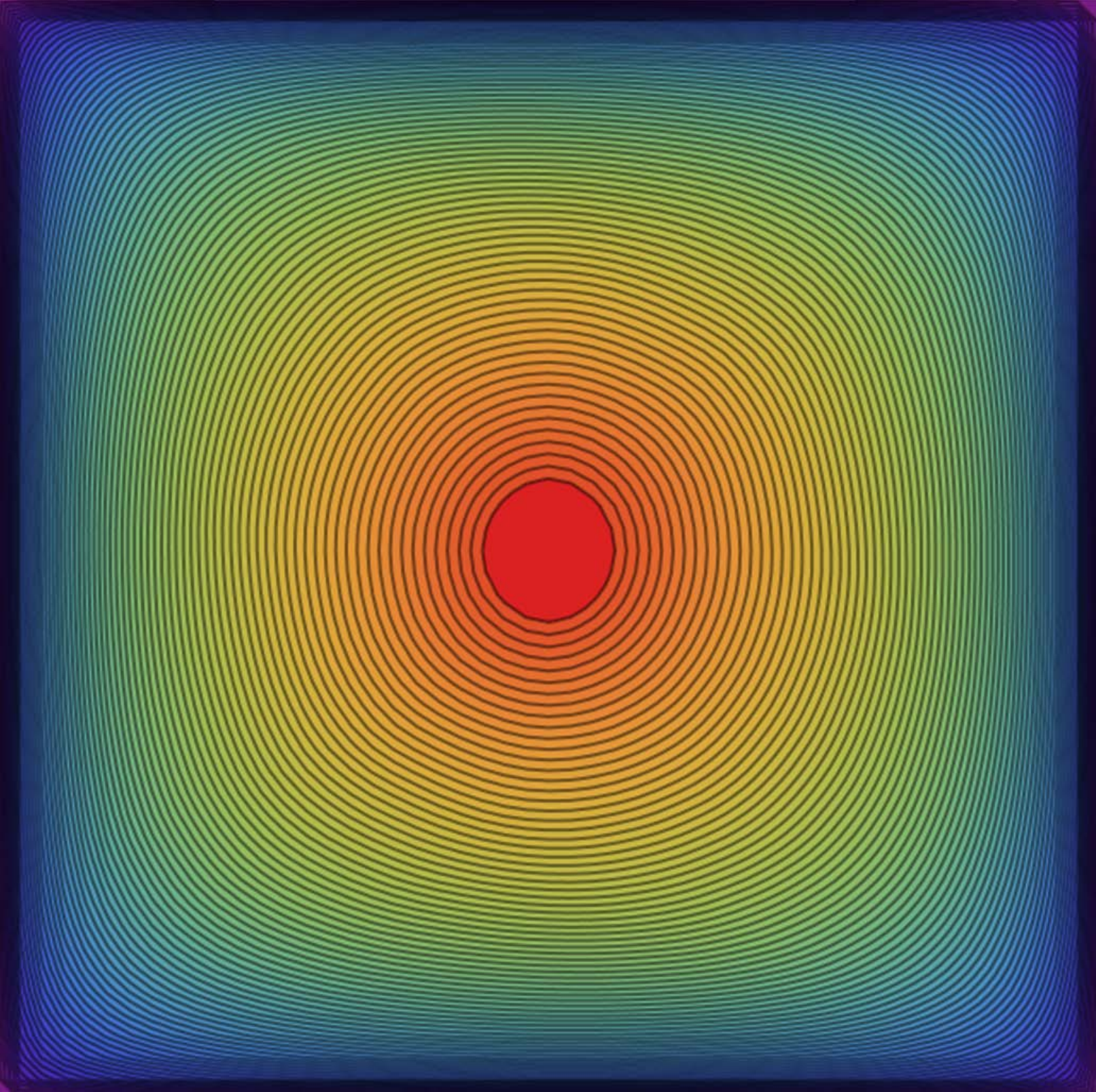}
 \caption{Level sets of the $h=1$ average on a square (generated by Ernesto Lupercio). Observe the gradual rounding of such equi-distant loci when the distance from the boundary grows. The small irregularities are due to Monte-Carlo approximation of the average by sampling random classes in $\operatorname{SL}_2(\mathbb{Z})\backslash\operatorname{SL}_2(\mathbb{R})$.}
 \label{fig:levels}
\end{figure}

At present, we do not know how to treat the maximal limit of equi-affine equi-distant loci in the compact case. However, for the analogous question, but in the non-compact case of a domain with two non-parallel asymptotes, it is relatively straightforward to show that the equi-affine equi-distant loci after a rescaling converge to branches of hyperbolas, which is the main proven result of this note, a first step toward a more general theory.

The prominence of ellipses and hyperbolae also relates to affine isoperimetric principles. In affine geometry \cite{blaschke1920affine, su1983affine}, ellipses often extremize or simplify affine-invariant functionals. For example, the affine isoperimetric inequality states that among all closed convex curves of a~given affine length, the ellipse encloses the minimal area. Similarly, the affine curvature flows tend to ``round out'' convex shapes into ellipses. In fact, affine curve shortening flow and affine normal flow provide a dynamic illustration: Sapiro and Tannenbaum \cite{sapiro1994affine} showed that evolving a~convex plane curve under motion by its affine normal causes the curve to become asymptotically ellipsoidal.

Explicit computations of values of the new class of equi-affine invariant functions using the definition written below is rather hard, as one needs to have an absolute control of tropical distance functions associated with all $\operatorname{SL}_2(\mathbb{R})$ transformations of the given domain. The only case for which our current knowledge is sufficient is the most symmetric of all, that is of the center of a disc, requiring understanding the deformations of singularities of tropical caustics over the space of ellipses -- the corresponding computation is carried out in Section~\ref{sec:example}.

\section{Definition, conjecture, theorem}

A tropical structure on $\mathbb{R}^2$ is a rank-two lattice \smash{$\Lambda \subset \bigl(\mathbb{R}^2\bigr)^*$}, viewed as a translation-invariant lattice of affine-linear slopes. Its co-area is the volume of the quotient torus \smash{$\bigl(\mathbb{R}^2\bigr)^*/\Lambda$}. The standard tropical structure is \smash{$\mathbb{Z}^2 \subset \bigl(\mathbb{R}^2\bigr)^*$}, which has co-area $1$.

The standard left action of $\operatorname{SL}_2(\mathbb{R})$ on $\mathbb{R}^2$ gives rise to the right action on \smash{$\bigl(\mathbb{R}^2\bigr)^*$} inducing a~transitive action on the set of tropical structures of co-area $1$. The stabilizer of the standard lattice is $\operatorname{SL}_2(\mathbb{Z})$. Therefore, this space is naturally identified with
$\operatorname{SL}_2(\mathbb{Z})\backslash \operatorname{SL}_2(\mathbb{R})$,
and carries the quotient Haar measure $\mu$.
Its total volume is finite and equals\footnote{More generally, for the measure induced by the Iwasawa decomposition ($\operatorname{SL}_n(\mathbb{R})=NAK$, where ${K=\operatorname{SO}(n)}$,~$A$~is Abelian, i.e., diagonal, and triangular $N$, exponents of nilpotent), its quotient by $\operatorname{SL}_n(\mathbb{Z})$ has volume $\zeta(2)\cdots\zeta(n)$, see the original article \cite{siegel}, or~\cite{garrett}.}
\[
\zeta(2)=\frac{\pi^2}{6}.
\]

Let $\Omega \subset \mathbb{R}^2$ be a convex domain, and let \smash{$\Lambda \subset \bigl(\mathbb{R}^2\bigr)^*$} be a tropical structure. Denote by
\[
\operatorname{rec}(\Omega):=\bigl\{v\in \mathbb{R}^2 \mid p+tv\in \Omega \text{ for all } p\in \Omega,\, t\ge 0\bigr\}
\]
the recession cone of $\Omega$.

For \smash{$\lambda \in \bigl(\mathbb{R}^2\bigr)^*$}, define
\[
c^\Omega_\lambda:=-\inf_{p\in \Omega}\lambda\cdot p \in \mathbb R\cup\{+\infty\}.
\]
Equivalently,
\[
c^\Omega_\lambda=h_\Omega(-\lambda),
\]
where $h_\Omega(\xi):=\sup_{p\in\Omega}\xi\cdot p$ is the support function of $\Omega$, possibly taking the value $+\infty$.

We say that $\Omega$ is $\Lambda$-admissible if there exists a nonzero $\lambda\in\Lambda$ such that $c^\Omega_\lambda<\infty$. Equivalently,~$\Omega$~is $\Lambda$-admissible if and only if there exists a nonzero $\lambda\in\Lambda$ such that
$\lambda(v)\ge 0$  for all $v\in \operatorname{rec}(\Omega)$.
Indeed, $c^\Omega_\lambda<\infty$ holds exactly when $\lambda$ is nonnegative on the recession cone of $\Omega$.

For a $\Lambda$-admissible convex domain $\Omega$, we define its tropical distance function relative to $\Lambda$ by
\[
\mathcal F^\Omega_\Lambda(p)
=
\inf_{\substack{\lambda\in\Lambda\setminus\{0\}\\ c^\Omega_\lambda<\infty}}
\bigl(c^\Omega_\lambda+\lambda\cdot p\bigr),
\qquad
p\in\Omega.
\]
Since $c^\Omega_\lambda+\lambda\cdot p\ge 0$ for every $p\in\Omega$, this defines a function
$
\mathcal F^\Omega_\Lambda\colon   \Omega\to [0,+\infty]$.

A lot can be said about such tropical series, see for instance \cite{kalinin2021shrinking, us_series, mikhalkin2023wave}. One important geometric property derived from the basic structure of Diophantine approximations is the following lemma.
\begin{Lemma}[{\cite[Lemma 4.4]{us_series}}]
For every $\Lambda$-admissible convex domain $\Omega$,
$\bigl(\mathcal F^\Omega_\Lambda\bigr)^{-1}(0)=\partial\Omega$.
\end{Lemma}
It is an important conceptual challenge to figure out if the tropical distance can be defined for a non-convex $\Omega$, see \cite{Shkolnikov2025PlanarTropicalCaustics}.

A convex domain $\Omega\subset\mathbb{R}^2$ is called strongly admissible if it is $\Lambda$-admissible for every tropical structure $\Lambda$ on the plane. Equivalently, $\Omega$ is strongly admissible if and only if it contains no~affine~lines.

\begin{Definition}
Let $\Omega\subset\mathbb{R}^2$ be a strongly admissible convex domain, and let $h>0$. We define the $h$-equi-affine distance function
$\mathcal{A}_{h}^\Omega\colon  \Omega\to[0,+\infty]$
by
\[
\mathcal{A}_{h}^\Omega(p)=
\biggl(
\frac{6}{\pi^2}
\int_{[\Lambda]\in \operatorname{SL}_2(\mathbb{Z})\backslash\operatorname{SL}_2(\mathbb{R})}
\bigl(\mathcal{F}^\Omega_\Lambda(p)\bigr)^{h}\,{\rm d}\mu[\Lambda]
\biggr)^{1/h}.
\]
In other words, $\mathcal{A}_{h}^\Omega(p)$ is the $h$-H\"older mean of the tropical distance functions over the space of tropical structures of co-area one.
\end{Definition}

The finiteness of $\mathcal{A}_{h}^\Omega$ is a nontrivial issue and will be discussed below. Whenever it is finite, the function $\mathcal{A}_{h}^\Omega$ is equi-affine invariant and homogeneous of degree $1$, that is,
\[
\mathcal{A}_{h}^{A(\Omega)}(A(p))=\mathcal{A}_{h}^{\Omega}(p)
\qquad\text{and}\qquad
\mathcal{A}_{h}^{r\Omega}(rp)=r \mathcal{A}_{h}^{\Omega}(p)
\]
for every $A\in\operatorname{SL}_2(\mathbb{R})$, every $p\in\Omega$, and every $r>0$.

One obvious property that allows to deduce the finiteness is the monotonicity of the values of the average with respect to inclusions, which is inherited from the corresponding monotonicity of the tropical distance function, following by its definition. Namely, if $p\in\Omega_1\subset\Omega_2$, then \[\mathcal{A}_{h}^{\Omega_1}(p)\leq\mathcal{A}_{h}^{\Omega_2}(p).\] As a corollary of Section \ref{sec:example}, for compact convex
$\Omega$ we obtain finiteness of $\mathcal A_h^\Omega(p)$ for every
$p\in\Omega$ and every $h>0$, by comparison with a disk centered at $p$
and containing $\Omega$.

We formulate the following conjecture.

\begin{Conjecture}
Let $\Omega\subset\mathbb{R}^2$ be a compact convex domain with nonempty interior, and let
\[
m_h^\Omega:=\max_{p\in\Omega}\mathcal A_h^\Omega(p).
\]
There exists $p_h$ the unique maximizer of $\mathcal A_h^\Omega$ and an ellipse $E_\Omega^h$ such that, as $t\uparrow m_h^\Omega$, the rescaled level sets \[\bigl(m_h^\Omega-t\bigr)^{-1}\bigl(\bigl(\mathcal A_h^\Omega\bigr)^{-1}(t)-p_h\bigr)+p_h\]
converge to $E_\Omega^h$ in the Hausdorff sense.
\end{Conjecture}

It is natural to ask whether the limiting ellipses $E_\Omega^h$ are in fact independent of $h$ up to rescalings and what their areas are. Due to the symmetry, one could expect that the resulting ellipse is proportional to Legendre's ellipse of the domain -- for instance for a square one gets a~circle, see Figure~\ref{fig:levels}. What the conjecture would give at the geometric level is an $\operatorname{SL}_2(\mathbb{R})$-invariant retraction to the space of ellipses of area one, since we know that the level sets of equi-affine distance function of the disk are circles due to rotation invariance.

In the next section, we prove the following unbounded analogue of the conjecture above for~${h\in(0,2)}$.

\begin{Theorem}
For $\Omega\subset\mathbb{R}^2$, a convex domain with a recession cone $Q=\operatorname{rec}(\Omega)$ bounded by two rays with support normal vectors $\lambda_1$ and $\lambda_2$ spanning a parallelogram of area one, $\mathcal{A}^\Omega_h(p+v_1)\leq \mathcal{A}^Q_h(p)=c_h\sqrt{\lambda_1(p)\lambda_2(p)}$ is finite for $h\in(0,2)$, where $v_1$ is the intersection of the two asymptotes of $\Omega$.

Moreover, for $h\in(0,2)$, the rescaled level sets
\[
t^{-1}\bigl(\mathcal A_h^\Omega\bigr)^{-1}(t)\longrightarrow H^\Omega_h=\bigl(\mathcal{A}^Q_h\bigr)^{-1}(1)
\]
converge to a branch of a hyperbola in the Hausdorff sense as $t\to+\infty$.
\end{Theorem}

\begin{Remark}
More geometrically, for every $h\in(0,2)$ there exists $c_h>0$ such that the limit of the rescaled level sets of $\mathcal{A}^\Omega_h$ is the branch of hyperbola with affine curvature \smash{$-\bigl(c_h^2/2\bigr)^{2/3}$} inscribed in the recession cone of every domain $\Omega$ with two non-parallel asymptotes. In other words, \smash{$-\bigl(c_h^2/2\bigr)^{2/3}$} is a~universal equi-affine quantity corresponding to the curvature of limiting hyperbolas for $h$-H\"older averaging over $\operatorname{SL}_2(\mathbb{Z})\backslash\operatorname{SL}_2(\mathbb{R})$ of tropical distance to the boundary.
\end{Remark}

The proof is based on a blow-down argument reducing the problem to the quadrant. In that model case, the averaged distance function is explicitly of the form $c_h\sqrt{xy}$, so its level sets are hyperbolas by the invariance under the diagonal subgroup of $\operatorname{SL}_2(\mathbb{R})$. Although we know they are finite for $h\in(0,2)$, we are unable to compute $c_h$ at the moment, as it would require a~much finer analysis of a combinatorial decomposition of $\operatorname{SL}_2(\mathbb{Z})\backslash\operatorname{SL}_2(\mathbb{R})$ induced by arithmetic interaction of cones with lattices.

\section{The proof}
Let $Q$ be the first quadrant, i.e., $Q=\bigl\{(x,y)\in\mathbb{R}^2 \mid x,y\geq 0\bigr\}$.
We show that \smash{$\mathcal{A}^Q_h(p)$} is finite for all $p\in Q$ if $h\in(0,2)$. Although the presented argument fails for $h\geq 2$, we expect a wider range of $h$ for the validity of finiteness of values.

Before passing to that, note that $Q$ is highly symmetric, namely it is invariant under homotheties with positive ratio and with respect to hyperbolic rotations \smash{$(x,y)\mapsto \bigl(sx,s^{-1}y\bigr)$}, where~${s>0}$~-- together, these two types of transformations act transitively on the interior of~$Q$. Thus, any function invariant under hyperbolic rotations and having homogeneous degree one with respect to homotheties is determined by its value $c$ at the point $(1,1)$, and has the form~$c\sqrt{xy}$. In~particular, \smash{$\mathcal{A}^Q_h$} is finite at every point if and only if \smash{$\mathcal{A}^Q_h(1,1)$} is finite.

The quantity \smash{$\mathcal{A}^Q_h(1,1)$} is the average
\[
\biggl(\frac{6}{\pi^2}\int_{\Lambda}(\min\{(a+b)\mid (a,b)\in\Lambda\backslash\{(0,0)\}; a,b\geq 0\})^h\, {\rm d}\mu(\Lambda)\biggr)^{1/h}
\]
over all lattices $\Lambda\subset\mathbb{R}^2$ such that the area of the torus $\mathbb{R}^2\slash\Lambda$ is equal to one. In~other words, instead of modifying $Q$ by $\operatorname{SL}_2(\mathbb{R})$ transformations, and keeping the standard tropical structure, we fix $Q$ and vary the tropical structures instead.

An estimate that we will be using is that \begin{equation}\tag{$\star$}\label{eq_star}
\min\{(a+b) \mid (a,b)\in\Lambda\backslash\{(0,0)\};\, a,b\geq 0\}<54(m(\Lambda))^{-1},\end{equation}
where $m(\Lambda)$ is the minimum of Euclidean lengths of all non-zero vectors in $\Lambda$. Then, the finiteness of $\mathcal{A}^Q_h(1,1)$ will follow from the convergence of the integral of $ m(\Lambda)^{-h}$ over $\Lambda\in\operatorname{SL}_2(\mathbb{Z})\backslash\operatorname{SL}_2(\mathbb{R})$ for $h<2$. The latter diverges for $h\geq 2$, though, and we would need to work harder at the level of interaction between number theory and tropical geometry to extend the finiteness of $\mathcal{A}^Q_h(1,1)$ for other $h$.

Fix a lattice $\Lambda$ of coarea one on the Euclidean plane. Let $e_1$ be a minimal length vector in~$\Lambda$, i.e., it has the length $|e_1|=m(\Lambda)$. Consider the set $E$ of all vectors $e\in\Lambda$ complementing $e_1$ to a~basis of $\Lambda$. The orthogonal projection of $E$ on the line $\mathbb{R}e_1$, identified with $\mathbb{R}$ by $s\mapsto s e_1$, forms an arithmetic progression with step~$m(\Lambda)$, thus its minimal positive element is at most~$m(\Lambda)$. Take $e_2$ to be the complementary basis vector to $e_1$ projecting on this element.

In other words, the component of $e_2$ parallel to $e_1$ has length at most $|e_1|$. On the other hand, the orthogonal to $e_1$ component of $e_2$ is $|e_1|^{-1}$ due to the coarea one condition on $\Lambda$. Together this gives an estimate on the length of $e_2$
\[|e_2|<\sqrt{|e_1|^2+|e_1|^{-2}}.\]

Since the open disk of radius $|e_1|$ centered at the origin does not contain non-zero points of $\Lambda$, Minkowski's theorem \cite{minkowski} gives that \smash{$|e_1|\leq \frac{2}{\sqrt{\pi}}$}, and we will use its cruder implication $|e_1|<2$. Taking square and dividing by $|e_1|$, we get \smash{$|e_1|<4|e_1|^{-1}$}, and so \smash{$|e_2|<\sqrt{17}|e_1|^{-1}<5|e_1|^{-1}$}.

The fundamental parallelogram spanned by $e_1$ and $e_2$ for the quotient $\mathbb{R}^2\slash\Lambda$ has the diameter at most $|e_1|+|e_2|$, which is less than $9|e_1|^{-1}$ by the estimates of the previous paragraph. In particular, for any point in $\mathbb{R}^2$, there exists a point of $\Lambda$ at distance at most $9|e_1|^{-1}$. Applying this to the point $\bigl(18|e_1|^{-1},18|e_1|^{-1}\bigr)$, there is a vector $(a,b)\in\Lambda$ such that $a,b>0$ and $a,b<27|e_1|^{-1}$. In particular, $a+b<54|e_1|^{-1}$, and so we arrive at (\ref{eq_star}).

Now we reduce the problem of showing $\mathcal{A}^Q_h(1,1)<+\infty$ to the convergence over $\Lambda\in\operatorname{SL}_2(\mathbb{Z})\backslash\operatorname{SL}_2(\mathbb{R})$ of the integral \[\int_\Lambda (m(\Lambda))^{-h}\, {\rm d}\mu(\Lambda).\]

Recall Siegel's mean value theorem \cite{siegel}. For any integrable $\phi\colon\mathbb{R}^2\rightarrow\mathbb{R}$,
\[
\frac{6}{\pi^2}\int_\Lambda\widehat\phi(\Lambda)\, {\rm d}\mu(\Lambda)=\int_{\mathbb{R}^2}\phi(x,y)\, {\rm d}x{\rm d}y,
\]
where \smash{$\widehat\phi$} is the Siegel transform of $\phi$ given by \smash{$\widehat\phi(\Lambda)=\sum_{v\in\Lambda\backslash(0,0)}\phi(v)$}.

Applying this to $\phi$ being the characteristic function of the disk of radius $\epsilon$ centered in $(0,0)$, we get \smash{$\mu\{\Lambda\mid m(\Lambda)<\varepsilon\}\leq \frac{\pi^3}{6}\epsilon^2$}. By the layer-cake formula,
\[
\int_\Lambda (m(\Lambda))^{-h}\,{\rm d}\mu(\Lambda)=h\int_{0}^\infty t^{h-1} \mu\bigl\{\Lambda\mid (m(\Lambda))^{-1}>t\bigr\} {\rm d}t.
\]

Again, as a corollary of Minkowski's theorem $(m(\Lambda))^{-1}>\frac{1}{2}$, and so
\[
h\int_{0}^\infty t^{h-1} \mu\bigl\{\Lambda\mid (m(\Lambda))^{-1}>t\bigr\} {\rm d}t=2^{-h}\frac{\pi^2}{6}+h\int_{\frac{1}{2}}^\infty t^{h-1} \mu\bigl\{\Lambda\mid (m(\Lambda))^{-1}>t\bigr\} {\rm d}t.
\]

Thus,{\samepage
\[\int_\Lambda (m(\Lambda))^{-h}\,{\rm d}\mu(\Lambda)\leq 2^{-h}\frac{\pi^2}{6}+\frac{\pi^3}{6}h\int_{\frac{1}{2}}^\infty t^{h-3} \,{\rm d}t,\]
which is finite for $h<2$.}

We are ready to prove the main theorem.

Let \smash{$\tilde\lambda_k(p)=\lambda_k(p-v_1)$}, where $k=1$ or $k=2$, and $v_1$ be the intersection point of the asymptotes of $\Omega$, and $\lambda_1$, $\lambda_2$ are linear functions represented by inner normal vectors to the recession cone $Q$ of $\Omega$ such that they span a parallelogram of area one. In particular, the equations~${\tilde\lambda_1(p)=0}$ and~${\tilde\lambda_2(p)=0}$ define the two asymptotes of $\Omega$. Let $Q_1=\bigl\{v\in\mathbb{R}^2\mid \tilde\lambda_1(v),\tilde\lambda_2(v)\geq 0\bigr\}=Q+v_1$. Since \smash{$\mathcal{A}^{Q_1}_h$} is finite everywhere on $Q_1$ for $h\in(0,2)$, and $\Omega\subset Q_1$, we have the finiteness of~$\mathcal{A}^{\Omega}_h$ on $\Omega$ by the monotonicity, and so we may speak about the non-empty level sets of~$\mathcal{A}^{\Omega}_h$. Next paragraph shows that \smash{$t^{-1}\bigl(\mathcal{A}_h^\Omega\bigr)^{-1}(t)$} Hausdorff converges to a branch of hyperbola $H$ as $t\rightarrow+\infty$, where the branch of a hyperbola is given by $\lambda_1(p)\lambda_2(p)=c_h^{-2}$ and~${\lambda_1(p),\lambda_2(p)>0}$.

Consider some point $v_2\in\Omega$ and denote by $Q_2$ the translation of the recession cone $Q$ of $\Omega$ by $v_2$. We have $Q_2\subset\Omega$. Since $Q$ with support functions $\lambda_1$ and $\lambda_2$ is $\operatorname{SL}_2(\mathbb{R})$-equivalent to the first quadrant \smash{$\bigl\{(x,y)\in\mathbb{R}^2\mid x,y\geq 0\bigr\}$} with support functions $x$ and $y$, where the $h$-average is equal to $c_h\sqrt{xy}$, we have
\[
\mathcal{A}^{Q_1}_h(v)=c_h\sqrt{\lambda_1(v-v_1)\lambda_2(v-v_1)}
\]
and
\[
\mathcal{A}^{Q_2}_h(v)=c_h\sqrt{\lambda_1(v-v_2)\lambda_2(v-v_2)}
\]
and so their rescaled by $t^{-1}$ level sets at $t$ are branches of hyperbolas
\[
H_{1,t}=\bigl\{p\in t^{-1}Q_1\mid \lambda_1\bigl(p-t^{-1}v_1\bigr)\lambda_2\bigl(p-t^{-1}v_1\bigr)=c_h^{-2}\bigr\}
\]
and
\[
H_{2,t}=\bigl\{p\in t^{-1}Q_2\mid \lambda_1\bigl(p-t^{-1}v_2\bigr)\lambda_2\bigl(p-t^{-1}v_2\bigr)=c_h^{-2}\bigr\},
\]
which differ by shift by $t^{-1}v_1$ and $t^{-1}v_2$ of the branch of hyperbola $H$, and so the Hausdorff distance between them and $H$ is at most $t^{-1}|v_1|_2$ and $t^{-1}|v_2|_2$ respectively. Since the rescaled level set \smash{$t^{-1}\bigl(\mathcal{A}^\Omega_h\bigr)^{-1}(t)$}, due to inclusion-monotonicity of equi-affine distance functions, is located between $H_{1,t}$ and $H_{2,t}$, its Hausdorff distance to $H$ tends to zero, and the Theorem follows.

\section{Example: Center of the disk}\label{sec:example}
In this section, we would like to carry out a concrete and explicit computation for a non-trivial value of some of the functions under consideration in the compact case. Take the standard unit disk $\bigcirc\subset\mathbb{R}^2$ as the underlying convex domain, we compute below $\mathcal{A}_1^\bigcirc$ at the center of the disk (other $h$ are fully analogous).

Again, we may think of $\mathcal{A}_1^{\bigcirc}(p)$ as the average of \smash{$\mathcal{F}_\Lambda^\bigcirc(p)$} over all tropical structures $\Lambda$, or as an average of \smash{$\mathcal{F}_{\mathbb{Z}^2}^{A\bigcirc}(Ap)$} over classes $[A]\in\SL_2(\mathbb{Z})\backslash\SL_2(\mathbb{R})$, i.e., fixing the standard tropical structure and varying within the orbit of the domain. The stabilizer of the disk $\bigcirc$ under $\SL_2(\mathbb{R})$ is precisely the rotation group $\operatorname{SO}(2)$, thus the orbit is identified with the upper half-space $\mathbb{H}$.

Under this identification, we think of points of $\mathbb{H}$ as of ellipses of area $\pi$ and with the origin as the center of symmetry. Such an ellipse is given by $q(x,y)=1$, where $q$ is the unique quadratic form $q(x,y)=ax^2+bxy+cy^2$ satisfying $a,c>0$ and $4ac-b^2=4$. Explicitly, the complex point in $\mathbb{H}$ is represented by the positive imaginary part root $x$ of the equation $q(x,1)=0$, which is~${-\frac{b}{2a}+\frac{{\rm i}}{a}}$.

To compute the tropical distance function (for the standard tropical structure) at an interior point of this ellipse, we need to figure out which of the tropical monomials minimizes the infinite tropical sum, i.e., the infimum in the definition. In general and by itself, this is a hard linear programming problem, however in the specific case of the center of the ellipse $q(x,y)=1$, the solution is easy to formulate, and we do below something slightly more subtle.

For each compact convex domain $\Omega$, we may consider the locus of points $p\in\Omega$ where $\mathcal{F}_\Lambda^\Omega$ attains its maximal value. Look at the set of tropical monomials contributing to $\mathcal{F}_\Lambda^\Omega$ near this locus. This information, i.e., the set of monomials $\lambda$ viewed as points in \smash{$\bigl(\ZZ^2\bigr)^*$}, together with its subdivision induced by coefficients $c_\lambda$, encodes the singularity type of the tropical caustic of~$\Omega$. Domains with the same caustic singularity type define a stratum in the space of all convex domains~\cite{thesism}. The miracle is that the restriction of this stratification to the space $\mathbb{H}$ of area $\pi$ ellipses coincides with the most standard stratification given by the translation of the interior $U=\bigl\{z\in\mathbb{C}\mid |{\operatorname{Re} z}|<\frac{1}{2},\, |z|>1\bigr\}$ of the fundamental domain of the $\SL_2(\mathbb{Z})$-action, as well as its edges and the boundary.

More specifically, ellipses corresponding to points $z\in U$ are exactly those where the monomials contributing to the minimum have gradients $(0, \pm 1)$, (and $(\pm 1,0)$ in case of a circle) and the tropical caustic has the vertical weight two edge (which passes through the origin). Thus, the value of the tropical distance series at $(0,0)$ is just $c_{(0,1)}$ from the monomial $c_{(0,1)}+y$ in the tropical series of the ellipse. In general, for the ellipse $q=1$ of area $\pi$ the $c_\lambda$ in its tropical distance series is given by $\sqrt{q^*(\lambda)}$, where $q^*$ denotes the dual quadratic form for which the $xy$ coefficient is replaced by its opposite (and $a$ and $c$ are switched) -- this can be deduced either from Legendre duality or Lagrange multipliers. Therefore, in the above notation, the value of the tropical distance series with standard tropical structure is~$\sqrt{a}$, i.e., the root of $x^2$ coefficient.\footnote{The language of tropical caustics can be avoided if we care only about the value at the center of the disc, and the only thing we actually need is to evaluate the tropical distance function of an ellipse $ax^2+bxy+cy^2=1$. The claim, which can be established by a direct comparison of tropical monomials, is that if this ellipse corresponds to a~point of the fundamental domain $U$, the value at the origin of the tropical distance is $\min_{(p,q)\in \mathbb{Z}^2\backslash\{(0,0)\}}\smash{\sqrt{q^*(p,q)}}=\smash{\sqrt{q^*(0,1)}=\sqrt{a}}$.}

With all the above, we compute the desired value as
\[
\mathcal{A}_1^\bigcirc(0,0)
=
(\operatorname{Area}_{\mathrm{hyp}} U)^{-1}
\int_U
(\operatorname{Im} z)^{-1/2}
\frac{{\rm d}\operatorname{Re} z\, {\rm d}\operatorname{Im} z}{(\operatorname{Im} z)^2}
=
\frac{4}{\pi}
\int_0^{\pi/6}
(\cos t)^{-1/2}\,{\rm d}t.
\]

Additionally, since $\operatorname{Im} z$ is bounded from below and $\operatorname{Re} z$ is bounded on $U$, the finiteness of~$\mathcal{A}_h^\bigcirc(0,0)$ is equivalent to the convergence of the integral
\[
\int_1^{\infty} y^{-\frac{h}{2}-2} {\rm d}_y,
\] which holds for $h>-2$. Interestingly, for the limiting case $h\rightarrow 0$ of the geometric mean the result is still valid. Together with inclusion-monotonicity, this gives the following.

\begin{Proposition}
For all $h>0$ and compact $\Omega$, the function $\mathcal{A}_h^\Omega$ is finite on $\Omega$. Moreover, the value of the arithmetic mean at the center of the unit disc is
\[\mathcal{A}_1^\bigcirc(0,0)=
\frac{4}{\pi}
\int_0^{\pi/6}
(\cos t)^{-1/2}\,{\rm d}t
\approx 0.6826794976.
\]
\end{Proposition}

\section{Discussion}
As classical geometry teaches us, it is customary, convenient, and easy to apply the results of a~more general affine geometry, with a larger group of symmetries, to a more specific geometry, such as Euclidean, with a smaller group of symmetries $\mathbb{R}^2\rtimes \operatorname{SO}(2)$. What happens in the present note is going backwards -- from tropical, with ``rotation'' group $\operatorname{SL}_2(\mathbb{Z})$, to equi-affine plane having larger non-translational part $\operatorname{SL}_2(\mathbb{R})$ of its symmetry group.

In addition, this is the second instance of an interaction between the tropical and the affine -- the other one is the $s=\frac{2}{3}$ zeta \smash{$s\mapsto\int_{p\in\Omega}\bigl(\mathcal{F}^\Omega_{\mathbb{Z}^2}(p)\bigr)^{s-2}{\rm d}p$} residue formula for the affine arc-length, a subject of a recent work~\cite{us_zeta}. In fact, it is conceivable that the equi-affine distance function would allow for a more transparent proof of that result.

There is yet another relation with another stream of our work, where the scaling limit of many-point relaxation in tropical sandpile is the main focus, see~\cite{KLSS26}. What happens there is that after rescaling by the square root of the number of perturbation points, the convergence of the corresponding tropical series to the solution of the Monge--Amp\'ere equation is established both theoretically and numerically, provided that the support of the perturbation density is isolated from the boundary. This in particular provides a different mechanism for the emergence of equi-affine invariance in tropical geometry. In addition, this motivates asking for the Monge--Amp\'ere measure of the averaged functions of the present article -- we conjecture that it is the sum of Lebesgue measure and of the Dirac measure supported at the center of mass of a given domain, with weights depending on the averaging parameter~$h$.

One stark difference between the affine and the tropical distance functions is that going to a level set of the tropical one defines a flow on the space of convex domains, the tropical wave front propagation, satisfying the most fundamental Huygens' principle, see \cite{mikhalkin2023wave}. It is unclear at the moment if the analogous property holds for equi-affine level sets at least for some averaging protocol, thus we refrain for the moment from using the terminology ``affine wave front''. Clarifying this is a crucial direction for future research.

We observe smoothness of non-zero level sets, and would not be surprised if they are analytic (and thus likely remember the boundary in contrast with the toric-blow-down information loss in the tropical wave front). Of course, by symmetry, we know that in the case of ellipses and hyperbolas, their level sets are again ellipses or hyperbolas. Similarly, a parabola also has a~one-parametric group of symmetries which is a subgroup of the semi-direct product of $\operatorname{SL}_2(\mathbb{R})$ and the group of translations~$\mathbb{R}^2$. The orbits of this subgroup under the action are parabolas. Therefore, the equi-affine equi-distant loci of all strictly convex conic domains are conic as well.

The phenomenon that maximal limits of equi-distant sets become conics, that is partially established above in the unbounded case, has an important implication in the compact case: the maxima of the equi-affine distance functions on compact domains would be attained in single points (this is another discrepancy with the tropical case, where the maximum is often attained along a segment). One may expect that these points are centers of mass, which might be deducible from the global topology of the space of convex domains modulo the respective group of symmetries.

It is important to note that the existence of $\mathcal{A}_h^\Omega$ relies on two fundamental properties. The first is that the space of tropical structures of fixed co-area has finite volume, which formally allows the averaging (for instance, the space of Euclidean structures with fixed area form -- $\H=\SL_2(\mathbb{R})/\operatorname{SO}(2)$ -- does not have this property). Second, the considered quantity must behave well with respect to the variation of tropical structure, i.e., be continuous or at least measurable, which would not hold for such basic invariants as point-to-point distances or angle's tropical cotangent, see \cite{mikhalkin2023wave}. Even so, the integral still may diverge -- the computation of the previous section implies that it does not happen for positive $h$ and compact $\Omega$.

\subsection*{Acknowledgements}
The authors are grateful to the anonymous referee whose comments and suggestions helped to significantly improve the paper.
M.S. would like to acknowledge the hospitality of GTIIT during the visit where this work was refined, and to thank Higinio Garcia Serrano for extremely valuable comments and the crucial idea of how the explicit computation of the arithmetic mean at the center of the unit disk can be simplified, as well Dmitrii Korshunov for numerous discussions, and pointing out, in particular, that the conjectural limit of the equi-distant loci of compact domain is may be universally proportional (for a given $h$) to a canonical covariant ellipsoid such as John's or Legendre's ellipsoid of the domain (including in higher dimensions). Work of M.S. is supported by the Simons Foundation, grant SFI-MPS-T-Institutes-00007697, and the Ministry of Education and Science of the Republic of Bulgaria, grant DO1-239/10.12.2024.

\pdfbookmark[1]{References}{ref}
\LastPageEnding

\end{document}